\documentstyle[preprint,aps]{revtex}
\tighten
\draft
\widetext
\input epsf
\preprint{YITP-SB-03-20}
\bigskip
\bigskip

\begin{document}

\title
    {DILUTING SOLUTIONS OF THE COSMOLOGICAL CONSTANT PROBLEM\footnote{B\lowercase{ased
 on talks given by \uppercase {A.I.} at \uppercase{N}ew \uppercase{Y}ork \uppercase{U}niversity,
\uppercase {O}hio \uppercase{S}tate \uppercase{U}niversity and the
\uppercase{SUGRA20} conference at \uppercase{N}ortheastern
\uppercase{U}niversity. \uppercase{A} shorter version of this
review will appear in the proceedings of the latter.}}}

\author{Olindo Corradini}
\address{Dipartimento di Fisica, Universit\`a di Bologna and INFN,
Sezione di Bologna\\
via Irnerio 46, I-40126 Bologna, Italy\\E-mail:
corradini@bo.infn.it}
\author{Alberto Iglesias}
\address{C.N. Yang Institute for Theoretical Physics\\
State University of New York, Stony Brook, NY 11794\\E-mail:
iglesias@insti.physics.sunysb.edu}
\author{Zurab Kakushadze}
\address{Royal Bank of Canada Dominion Securities
Corporation\footnote{This address
is used by the corresponding author for no purpose other than to indicate his
professional affiliation as is customary in scientific publications. In
particular, the contents of this paper are limited to Theoretical Physics, have
no relation whatsoever to the Securities Industry, do not have any value for
making investment or any other financial decisions, and in no way represent
views of the Royal Bank of Canada or any of its other affiliates.}\\
1 Liberty Plaza, 165 Broadway, New York, NY 10006\\
E-mail: zurab.kakushadze@rbccm.com}

\maketitle

\begin{abstract} 
We discuss the cosmological constant problem in the
context of higher codimension brane world scenarios with
infinite-volume extra dimensions.
\end{abstract}

\section{Introduction}

{}The cosmological constant problem is an outstanding obstacle in
the quantum field theory description of gravity interacting
with matter. The problem is to reconcile the expected contribution
of the zero point energy of matter and gauge fields to the vacuum
energy density with the observations (coming from Supernova and
Cosmic Microwave Background data), which imply that the universe is
flat with a positive cosmological constant. The former (assuming supersymmetry
breaking around TeV) is at least of order TeV$^4$, while the latter is at
most of order $(10^{-15}~{\rm TeV})^4$.

{}We will present a possible way to circumvent the problem by
allowing gravity to propagate in infinite volume extra dimensions
while the Standard Model fields are localized on a lower
dimensional hypersurface (brane).

{}In order to introduce some general features of brane world
scenarios with infinite volume extra dimensions and summarize some
results we can consider the following action
\begin{equation}
S={\widetilde M}_P^{D-d-2}\int_\Sigma d^{D-d}x~\sqrt{-{\widetilde
G}}\left({\widetilde R}-{\widetilde
\Lambda}\right)+M_P^{D-2}\int_{\mu_D}d^{D}x~
 \sqrt{-{G}}R~.
\end{equation}
Here $\mu_D$ is the D-dimensional bulk in which gravity
propagates, and the second term is a D-dimensional
Einstein-Hilbert term with a fundamental gravitational scale
$M_P$, while $\Sigma$ is a lower dimensional hypersurface, a
brane, with tension ${\widetilde M}_P^{D-d-2}{\widetilde\Lambda}$
(which includes contributions to the vacuum energy density from
the Standard Model fields whose dynamics is otherwise neglected).
The brane Einstein-Hilbert term containing ${\widetilde R}$ is
induced\footnote{Along with higher derivative terms as well as
nonlocal terms.} via loops of gauge and matter fields interacting
with bulk gravity as long as the world volume theory is
non-conformal. ${\widetilde M}_P$ is then related to a world
volume theory scale and is identified with $1/\sqrt{G_N}$. Lower
dimensional gravity is reproduced at distances ranging from
$1/M_P$ to ${\widetilde M}_P/M_P^2$ becoming higher dimensional at
larger distances, while below the $1/M_P$ scale the effective
field theory description breaks down. Gravity is localized on the
brane with only ultralight modes leaking into the bulk. This was
shown for the case of a tensionless brane case\cite{DG} ({\it
i.e.}, when ${\widetilde \Lambda}=0$), and for the case of a
non-zero tension 3-brane in 6 dimensional bulk (and, more
generally, for the case of a non-zero tension codimension-2
brane)\cite{CIKL}.

{}Within this context, a step toward solving the
 cosmological constant problem would then be to find solutions
 where the brane is flat without having to finetune the brane
 tension {\it i.e.}, for a continuous range of values of ${\widetilde\Lambda}$
 that
 would account for the zero point energies of the brane fields. We
 refer to such solutions as ``diluting''.
Such solution do not seem to exist
when only one extra dimension is considered,
therefore, we will focus on higher codimension setups.

{}Considering a higher codimension brane with no thickness (a
$\delta$-function like brane) leads to a singular graviton
propagator at the location of the brane. Taking $r$ to be the radial
direction in the extra space, it grows like ${\rm
log}(r)$ for the case of a brane with two extra dimensions, and as
$r^{2-d}$ for branes of codimension $d>2$.

{}In the tensionless case the extra space can be taken to be
Minkowski space. However, when ${\widetilde \Lambda}$ is
non-vanishing the extra space is no longer flat and a singularity
is already found in the background solution. In the codimension-2
case it is a wedge with a deficit angle and the singularity is a
conical one at the location of the brane. Here, a range of
tensions is consistent with a flat brane\footnote{A scale as low
as $10^{-15} {\rm TeV} \sim 0.1mm$ is required in these models, in
such a way that the modification of the laws of gravity appears at
scales, larger than the present-day Hubble scale ($10^{29} mm$)
and smaller than $0.1~mm$ \cite{DGHS}.}, namely\cite{CIKL},
\begin{equation}
0\leq T {\
\lower-1.2pt\vbox{\hbox{\rlap{$<$}\lower5pt\vbox{\hbox{$\sim$}}}}\
}4\pi M_P^4\sim (10^{-15}~{\rm TeV})^4~,
\end{equation}
which does not improve the observed bound. In this light we are
led to consider branes of codimension 3 or higher. The
singularity of the background is still an issue in these cases and,
furthermore, another type of singularity appears at a finite
distance from the brane\cite{Gregory}. The topic of the next
section is how to cure this singular behavior of such solutions.


\section{Singularities}

{}In order to cure the singular behavior at the origin we give
some structure to the brane in the extra space. Thus, instead of
taking the brane to be just a point in the extra dimensions we
could grow this point into a $d$-dimensional ball ${\bf B}_d$ of
non-zero radius $\epsilon$. This type of resolution raises a
problem, the gravitational modes on the brane contain an infinite
tower of tachyonic modes (already for the tensionless case\cite{orient}).
This can be circumvented by considering a partial
smoothing where one replaces the $d$-dimensional ball by a
$(d-1)$-dimensional sphere ${\bf S}^{d-1}$ of radius $\epsilon$.
This suffices to smooth out higher codimension singularities in the
propagator because close to the brane effectively we now have
a codimension-1 brane for which
the propagator is non-singular. Moreover, in the tensionless case
as well as in the case of a non-zero tension codimension-2 brane
we will then have only one tachyonic mode\footnote{In the case of
solitonic branes one does not expect tachyons at all \cite{KPR}. However, 
we are not considering here a solitonic brane, rather the original setup
of \cite{DG}.} (with ultra-low
$-m^2$) which is expected to be an artifact of not including non-local
operators on the brane. As to the background itself this procedure
does smooth out $r=0$ singularities. Thus, now we are
considering the following action
\begin{equation}
S={\widetilde M}^{D-3}_P\int_{\Sigma}d^{D-1}x~\sqrt{-{\widetilde G}}
 \left({\widetilde R}
-{\widetilde\Lambda}\right)+M^{D-2}_P\int_{{\mu}_D}d^{D}x~
 \sqrt{-{G}}R~,
\end{equation}
where $\Sigma={\bf R}^{D-d-1,1}\times {\bf S}^{d-1}$. However, not
surprisingly, this procedure does not suffice for smoothing out
the aforementioned finite-distance singularities that appear in
the codimension-3 and higher cases\cite{CIKL}. It was suggested
that higher curvature terms in the {\em bulk} might be responsible
for curing such singularities\cite{Higher}. Studying such
backgrounds in the presence of higher curvature terms in the bulk
is rather non-trivial and in order to make the problem more
tractable, we focus on a special kind of higher curvature terms in
the bulk. In particular, we consider adding the second-order (in
curvature) Gauss-Bonnet combination in the bulk. This combination
has some remarkable properties. Thus, it is an Euler invariant in
four dimensions (but it contributes nontrivially to the equations
of motion in higher dimensions). It appears in Supergravity
theories as the necessary combination to supersymmetrize the
Chern-Simons term, and is also expected to appear in a low energy
effective action derived from string theory as it is the only
combination that does not introduce propagating ghosts. But for
our purposes the most important property is  that it does not
introduce terms with third and fourth derivatives in the
background equations of motions, but rather makes them even more
non-linear. Albeit non-trivial, these equations can in certain
cases be analyzed analytically, so we can get some insight into
the effect of higher curvature terms on the finite distance
singularities. Our action now reads:
\begin{eqnarray}
 S=&&{\widetilde M}^{D-3}_P \int_{\Sigma} d^{D-1}x~\sqrt{-{\widetilde G}}\left[{\widetilde
R}-{\widetilde\Lambda}\right]+\nonumber\\
 &&{M}^{D-2}_P \int d^{D}x~
 \sqrt{-{G}}~\left[{R}~+\xi\biggl(R^2-4R_{MN}^2+R_{MNPR}^2\biggl)\right],
 \label{action}
\end{eqnarray}
where $\xi$ is the Gauss-Bonnet coupling.

{}We are interested in solutions with vanishing
$(D-d)$-dimensional cosmological constant, which, at the same
time, are radially symmetric in the extra $d$ dimensions. The
corresponding ansatz for the background metric reads:
\begin{equation}
 ds^2=\exp(2A)~\eta_{\mu\nu}~dx^\mu dx^\nu+\exp(2B)~\delta_{ij}~dx^i dx^j~,
\end{equation}
where $A$ and $B$ are functions of $r$ but are independent of
$x^\mu$ (the $(D-d)$ coordinates on the original brane) and $x^\alpha$
(the angular coordinates on the extra space over which the brane
was smoothed). The metric for the coordinates
$x^i=(x^\alpha,r)$ is conformally flat.

{}The equations of motion are highly
non-linear and difficult to solve in the general case. However, in
the $d=4$ and especially $d=3$ cases various (but not all) terms
proportional to $\xi$ vanish. This is due to the fact that the
Gauss-Bonnet combination is an Euler invariant in four dimensions.
To make our task more tractable, from now on we will focus on the
codimension-3 case ($d=3$). We do {\em not} expect higher
codimension cases to be qualitatively different.

{}Not only is the complexity of the equations of motion
sensitive to the value of $d$, but also to the value of
${\overline d}\equiv D-d$. In particular, we have substantial
simplifications in the cases of ${\overline d}=2$ and ${\overline
d}=3$ corresponding to the non-compact part of the brane $\Sigma$
being a string respectively a membrane. Note that these
simplifications are specific to the Gauss-Bonnet combination. In
particular, if we set the Gauss-Bonnet coupling $\xi$ to zero
(that is, if we keep only the Einstein-Hilbert term in the bulk),
there is nothing special about the ${\overline d}=2,3$ cases. This
suggest that the conclusions derived from explicit analytical
computations for the ${\overline d}=2,3$ cases can be expected to
hold in ${\overline d}\geq 4$ cases as well (in particular, in the
case of a 3-brane in 7D and, as we mentioned above, even higher
dimensional bulk). In fact, the analytical and numerical results
in the case of a 3-brane in 7D bulk confirm this expectation.

{}What follows is a review of the results of a recent
collaboration\cite{CIK}.

\section{No Einstein-Hilbert bulk term}

{}In order to test the hypothesis of smoothing out by the presence
of higher curvature terms we first studied a more of a toy model
type of action in which the bulk Einstein-Hilbert term was
neglected. Let us first consider a string ($\overline d=2$)
propagating in 5D bulk ($d=3$). The equations of motion then read:
\begin{eqnarray}\label{S-eq1}
&&(A^\prime)^2 \left[3(B^\prime)^2+6{B^\prime\over r}+
{2\over r^2}\right]=0~,\\
&& \left[A^{\prime\prime}+(A^\prime)^2-A^\prime B^\prime\right]
\left[(B^\prime)^2+2{B^\prime\over r}\right] +2A^\prime
\left(B^\prime +{1\over r}\right)\left(B^{\prime\prime}
+{B^\prime\over r}\right)=\nonumber\\
&&{1\over 8\xi}{\rm e}^{3B}(\kappa-2)\lambda\widetilde L
\delta(r-\epsilon)~,\label{S-eq2}\\
&&(A^\prime)^2\left(B^{\prime\prime}+{B^\prime\over r}\right)
+2\left[A^{\prime\prime}+(A^\prime)^2-A^\prime B^\prime\right]
A^\prime \left(B^\prime +{1\over r}\right)=\nonumber\\
&&{\kappa\over 8\xi}{\rm e}^{3B}\lambda\widetilde L
\delta(r-\epsilon)~,\label{S-eq3}
\end{eqnarray}
where we have introduced the notation
\begin{equation}
\widetilde L={\widetilde M_P^{D-3}\over M_P^{D-2}}~,~~~~~~
\lambda={d-2\over\epsilon^2}{\rm e}^{-2B}~,~~~~~~
\kappa\equiv{\widetilde \Lambda}/\lambda~.
\end{equation}
We solve these equations for $r\not =\epsilon$ and then impose the
two matching conditions for the $r<\epsilon$ and $r>\epsilon$
parts of the solution that produce the jump in the first
derivatives of the warp factors implied by the $\delta$ functions
on the r.h.s. of (\ref{S-eq2}) and (\ref{S-eq3}).

{} For {$r\not =\epsilon$} flat space is a solution, {\it i.e.}, we can
simultaneously set
$A$ and $B$ to be constant. A nontrivial solution is also easily found,
namely,
\begin{equation}\label{S-nontr}
A(r)={\rm ln}\left({r_A\over r}\right)^{\alpha -1},~~~ B(r)={\rm
ln}\left({r_B\over r}\right)^\alpha~,
\end{equation}
where
\begin{equation}
 \alpha=\alpha_\pm = 1\pm {1\over \sqrt 3}~.
\end{equation}
Thus we can consider two possibilities: Either we
take the solution to be nontrivial inside the spherical shell at
$r=\epsilon$, while outside we take constant warp factors; or we
consider flat space inside the sphere taking the nontrivial part
of the solution on the outside. We will refer to the former
solutions as {\it interior} and call the latter solutions {\it
exterior}.

{}Let us analyze the singularity structure in the above solutions.
Singularities can potentially occur at $r\rightarrow \infty$ in
the exterior solutions and at $r\rightarrow 0$ in the interior
ones. Thus, the line element in the non-trivially warped part of
the space-time is given by:
\begin{equation}
ds^2=  \left({r_A\over r}\right)^{2(\alpha-1)} \eta_{\mu\nu}dx^\mu
dx^\nu +\left({r_B\over r}\right)^{2\alpha} \delta_{ij} dx^i dx^j
~,
\end{equation}
which is singular at $r=0$ for both roots $\alpha=\alpha_\pm$.
However, only for $\alpha=\alpha_-$ is the space truly singular,
whereas for $\alpha=\alpha_+$ we merely have a coordinate
singularity. Thus, consider a $2n$-derivative scalar. Such an
object - let us call it $\chi^{(n)}$ - has the following
expression in terms of the extra-space radius
\begin{equation}\label{n-scalar}
\chi^{(n)}\sim {\rm e}^{-2n B} {1\over r^{2n}} \sim r^{2n(\alpha
-1)}.
\end{equation}
The latter blows up for $\alpha=\alpha_-$ as $r$ approaches zero;
this singularity is a naked singularity. One can indeed consider
radial null geodesics with affine parameter $\sigma$ and use that
$G_{tt}dt/d\sigma$ is constant along geodesics to obtain
\begin{equation}
{dr\over d\sigma} \sim r^{2\alpha-1}~.
\end{equation}
For $\alpha=\alpha_-$ these geodesics terminate with finite affine
parameter as
 $r$ approaches zero:
\begin{equation}
\sigma \sim r^{2(1-\alpha)}+{\rm constant}~.
\end{equation}
Thus, we have incomplete geodesics reaching a point of divergent
curvature. On the other hand, for $\alpha=\alpha_+$ the expression
(\ref{n-scalar}) vanishes as $r$ approaches zero, the
aforementioned geodesics are complete ({\it i.e.},
$\sigma\rightarrow\infty$), and it is not difficult to see, by
doing a similar calculation, that radial time-like geodesics
extend to infinite proper time in this limit. Therefore, $r=0$ is
a coordinate singularity in this case. This then implies that for
the interior solutions we must choose $\alpha=\alpha_+$. It then
follows that the Gauss-Bonnet coupling $\xi$ is {\em positive} in
this case.

{}Similar considerations apply to the $r\rightarrow\infty$
singularity. In this case we have a naked singularity for
$\alpha=\alpha_+$, while for $\alpha=\alpha_-$ we merely have a
coordinate singularity. This implies that for the exterior
solutions we must choose $\alpha=\alpha_-$. Note that in this case
the Gauss-Bonnet coupling $\xi$ is also {\em positive}. Moreover,
it is also not difficult to check that the volume of extra space
is infinite.

{}Thus, as we see, we have sensible infinite-volume {\em
non-singular} solutions if we take the bulk action to be given by
the Gauss-Bonnet combination. However, these solutions exist only
for a fine-tuned value of the brane tension. Indeed,
the matching conditions at $r=\epsilon$ for the interior ($+$)
and exterior ($-$) solutions reduce to:
\begin{eqnarray}
&& \pm {\alpha(\alpha-1)(\alpha-2)\over \epsilon}=
 -{\kappa -2\over 8\xi}{\widetilde L}
\left({r_B\over \epsilon}\right)^\alpha~,\\
&&\pm {(\alpha-1)^3\over \epsilon} =-{\kappa\over 8\xi}{\widetilde
L} \left({r_B\over \epsilon}\right)^\alpha~,
\end{eqnarray}
Note that the brane tension is always positive (assuming that the
``brane-width'' $\epsilon$ is non-zero). An unwelcome feature of
these solutions is that the tension is determined by the other
parameters (the integration constant $r_B$ drops out): which
implies that
\begin{equation}\label{CC-string}
 {\widetilde \Lambda}={8{\widetilde L}^2\over\xi^2}~.
\end{equation}
That is, these solutions are {\em not} ``diluting'', rather they
exist only if we fine-tune the brane tension and the Gauss-Bonnet
coupling. In the following we will see that this is specific to the
case at hand, and diluting solutions do exist in other cases. In
all the other cases we studied, for which the equations are not so
simple, a dependence on one of the integration constants from the
nontrivial part of the solution appears in the l.h.s. of the
matching conditions. This extra freedom translates into a brane
tension being unfixed by the other parameters of the theory.

{}The case of a membrane in 6D bulk can also be solved
analytically. We found smooth {\it diluting} interior as well as exterior
solutions in this case. In the case of a 3-brane in 7D
bulk we found smooth (with just a coordinate singularity at $r=0$) {\it
diluting} solutions with positive brane tension via
numerical methods.

{}Encouraged by these results we then studied the case in which
both the bulk Einstein-Hilbert and Gauss-Bonnet terms are present.


\section{Einstein-Hilbert and Gauss-Bonnet terms in the bulk}

{}Let us now concentrate on the case of a 3-brane in 7-dimensional
space with both Einstein-Hilbert and Gauss-Bonnet bulk terms, which
is the most interesting case from the phenomenological point of
view. By introducing the following variables:
\begin{equation}
V=rA^\prime~,\hskip1cm U=rB^\prime+1~,\hskip1cm z={r^2{\rm
e}^{2B}\over\xi}~,
\end{equation}
the ($rr$) equation can be rewritten as
\begin{equation}\label{EH3b4}
U^2(72V^2-z)-2U4V(z-12V^2)-z(6V^2-1)-12V^2(2-V^2)=0~,
\end{equation}
which can be used to express $U$ as a function of $V$ and $z$. And
similarly, we can rewrite the ($rr-\alpha\beta$) equation of
motion which gives, away from the brane, the following first order
differential equation for $V$ (treating $V$ as a function of $z$):
\begin{equation}\label{dvdz}
2zU{dV\over dz}={g_1\over g_2}~,
\end{equation}
where $g_1$ and $g_2$ are known polynomials of $U$, $V$ and $z$. And,
as in the previous cases, the solution must satisfy two matching
conditions at $r=\epsilon$.

{}In order to have non-trivial solutions consistent with the
matching conditions, $V(\epsilon+)$ (for exterior solutions),
$V(\epsilon-)$ (for interior solutions) and $z(\epsilon)$ must be
chosen in certain regions of the $V-z$ plane.

{}For the positive Gauss-Bonnet coupling case, $\xi>0$ ({\it
i.e.}, $z>0$) some of the interior solutions that are consistent
with the matching conditions are smooth with just a
coordinate singularity $r\rightarrow r_0+$ limit (where $r_0$ is
an integration constant). Both the affine
parameter of radial null
geodesics and the proper time of time-like geodesics diverge in
that limit as $1/{\rm ln}(r/r_0)$. Thus, space is complete if
we cut it at $r=r_0$. The values of $\kappa$ consistent with the
matching conditions range from $-\infty$ to 2.15. These solutions
are of particular interest as some of these solutions have positive
3-brane tension (the ones with $\kappa>2$) and, furthermore, they
are diluting.

{}In order to make contact with the phenomenology mentioned in the
introduction we have to give a more precise definition of the
brane tension. Up to this point we have been referring to
${\widetilde\Lambda}$ (or, more precisely, ${\widetilde T}\equiv
{\widetilde M}_P^{D-3}{\widetilde\Lambda}$) as the brane tension.
This quantity is indeed the tension of the smoothed brane $\Sigma$
whose world-volume has the geometry of ${\bf R}^{{\overline
d}-1,1}\times {\bf S}^{d-1}_\epsilon$. Note, however, that a bulk
observer at $r>\epsilon$ does not see this brane tension - to such
an observer the brane appears to be tensionless. Indeed, the warp
factors are constant in the aforementioned solution at
$r>\epsilon$. The non-vanishing (in fact, positive) tension of the
brane $\Sigma$ does not curve the space outside of the sphere
${\bf S}^{d-1}_\epsilon$. Instead, it curves the space {\em
inside} the sphere ${\bf S}^{d-1}_\epsilon$, that is, at
$r<\epsilon$. And this happens without producing any singularity
at $r<\epsilon$, and with the non-compact part of the world-volume
of the brane remaining flat.

{}Here it is important to note that the effective tension of the
{\em fat} 3-brane (in the present case) whose world-volume is
${\bf R}^{3,1}$ (this 3-brane is fat as it is extended in the
extra 2 angular dimensions) is also positive. It is not difficult
to see that this brane tension is given by
\begin{equation}
 {\widehat T}=({\widetilde\Lambda}-2\lambda)v_2
 {\widetilde M}_P^4~,
\end{equation}
where $v_2=4\pi\epsilon^2 e^{2B(\epsilon)}$ is the volume of the
sphere ${\bf S}^2_\epsilon$ whose radius is not $\epsilon$ but
rather
\begin{equation}
 R\equiv\epsilon e^{B(\epsilon)}~.
\end{equation}
We therefore have:
\begin{equation}
 {\widehat T}=4\pi(\kappa-2)
 {\widetilde M}_P^4~.
\end{equation}
Thus, for $\kappa>2$ (which is part of the parameter space for the
aforementioned solution) this effective fat brane tension is {\em
positive}.

{}Next, the 4-dimensional Planck scale ${\widehat M}_P$ is given
by
\begin{equation}
 {\widehat M}_P^2=v_2 {\widetilde M}_P^4=4\pi R^2 {\widetilde M}_P^4~.
\end{equation}
where ${\widehat M}_P$ is the 4-dimensional Planck scale,
${\widetilde M}_P$ is the 6-dimensional Planck scale.

{}{\em A priori} we can reproduce the 4-dimensional Planck scale
${\widehat M}_P\sim 10^{18}~{\rm GeV}$ by choosing $R$ between
$R\sim {\rm millimeter}$ and $R\sim 1/{\widehat M}_P$. The
6-dimensional Planck scale then ranges between\footnote{Note a
similarity with \cite{ADD}.} ${\widetilde M}_P\sim {\rm TeV}$ and
${\widetilde M}_P\sim {\widehat M}_P$. On the other hand, it is
reasonable to assume a ``see-saw'' modification of gravity\cite{DGHS}
in the present case,
in particular, once we take into account higher curvature terms on
the {\em brane}, to obtain 4-dimensional gravity up to the
distance scales of order of the present Hubble size, we must
assume that the ``fundamental'' 7-dimensional Planck scale
\begin{equation}
 M_P\sim ({\rm millimeter})^{-1}~.
\end{equation}
Let us see what range of values we can expect for the 5-brane
tension ${\widetilde T}$.

{}Thus, the 5-brane tension
\begin{equation}
 {\widetilde T}=
 {\widetilde \Lambda}{\widetilde M}_P^4=\kappa R^{-2}{\widetilde M}_P^4
 \sim R^{-4} {\widehat M}_P^2~.
\end{equation}
If we assume that the Standard Model fields come from a
6-dimensional 5-brane theory compactified on the 2-sphere, then we
might need to require that $R{\
\lower-1.2pt\vbox{\hbox{\rlap{$<$}\lower5pt\vbox{\hbox{$\sim$}}}}\
} ({\rm TeV})^{-1}$. Then ${\widetilde M}_P$ ranges between
$10^7-10^8~{\rm TeV}$ and ${\widehat M}_P$, while the 5-brane
tension ranges between $(10^{5}~{\rm TeV})^6$ and ${\widehat
M}_P^6$. Note that {\em a priori} this is not in conflict with
having the supersymmetry breaking scale in the TeV range.

{}In principle, the above scenario {\em a priori} does not seem to
be inconsistent modulo the fact that we still need to explain why
the 6-dimensional Plank scale ${\widetilde M}_P$ is many orders of
magnitude (30 in the extreme case where $R^{-1}\sim{\widetilde
M}_P\sim {\widehat M}_P$) higher than the seven-dimensional Planck
scale $M_P$. Note, however, that the same issue is present in any
theory with infinite-volume extra dimensions.

{}Let us see what kind of values of $V(\epsilon-)$ we would need
to have in order to obtain a solution satisfying the above
phenomenological considerations. First, we will assume that the
Gauss-Bonnet parameter $\xi\sim M_P^{-2}$ (its ``natural'' value).

{}We can for instance take the $z\ll |V|$ ($|V|\gg 1$) case in
which the matching conditions become,
\begin{eqnarray}
R{\widetilde L}/\xi&\approx& -11.77V^3(\epsilon-)~,\\
\kappa R{\widetilde L}/\xi&\approx& -25.3V^3(\epsilon-)~,
\end{eqnarray}
which gives $\kappa\approx 2.15$, and $|V(\epsilon-)|\approx
10^{30}$, and $\epsilon$ is very close to the would-be singularity
$r_*$ (here for definiteness we have assumed the extreme case
$R^{-1}\sim {\widetilde M}_P\sim {\widehat M}_P$, where
${\widetilde L}={\widetilde M}_P^4/ M_P^5\sim 10^{120}~{\rm mm}$):
\begin{equation}
 {r_*\over \epsilon}-1\sim 10^{-30}~.
\end{equation}
Note that the singularity at $r=r_*>\epsilon$ would be there if we
took the interior solution for $r<\epsilon$ and continued it for
values $r>\epsilon$. However, in our solution (just as in the case
without the Einstein-Hilbert term) there is no singularity as for
$r>\epsilon$ the warp factors are actually constant, and this is
consistent with the matching conditions at $r=\epsilon$. Thus, as
we see, in the
presence of both the Einstein-Hilbert and Gauss-Bonnet terms in
the bulk action we have sensible smooth solutions with positive
brane tension. Moreover, these solutions are {\em diluting}, that
is, they exist for a range of values of the brane tension (note
that the Gauss-Bonnet coupling in these solutions is {\em
positive}). In particular, no fine tuning appears to be required in
our solution.

{}As we mentioned above, the singularity at finite
$r=r_0<\epsilon$ is harmless as the corresponding geodesics are
complete. Note that in the case without the Einstein-Hilbert term
the corresponding coordinate singularity is at $r=0$. The reason
why is the following. If we start with a solution corresponding to
both the Einstein-Hilbert and Gauss-Bonnet terms present in the
bulk, to arrive at the solution corresponding to only the
Gauss-Bonnet term present in the bulk we must take the limit
$\xi\rightarrow\infty$, $M_P^{D-2}\xi={\rm fixed}$. It is then not
difficult to check that in this limit the coordinate singularity
at $r=r_0$ continuously moves to $r=0$. Also note that, since the
singularity at $r=r_0$ in the full solution is a coordinate
singularity, we can consistently cut the space at $r=r_0$. The
geometry of the resulting solution then is as follows. In the
extra three dimensions we have a radially symmetric solution where
a 2-sphere is fibered over a semi-infinite line $[r_0,\infty)$.
The space is curved for $r_0\leq r<\epsilon$, at $r=\epsilon$ we
have a jump in the radial derivatives of the warp factors (because
$r=\epsilon$ is where the brane tension is localized), and for
$r>\epsilon$ the space is actually {\em flat}. So an outside
observer located at $r>\epsilon$ thinks that the brane is
tensionless, while an observer inside of the sphere, that is, at
$r<\epsilon$ sees the space highly curved (and it would take this
observer infinite time to reach the coordinate singularity at
$r=r_0$). This is an important point, in particular, note that we
did not find smooth exterior solutions where the space would be
curved outside but flat inside. And this is just as well. Indeed,
if we have only the Einstein-Hilbert term in the bulk, then we
have no smooth solutions whatsoever\cite{Higher} (that is,
smoothing out the 3-brane by making it into a 5-brane with two
dimensions curled up into a 2-sphere does not suffice). What is
different in our solutions is that we have added higher curvature
terms in the bulk, which we expected to smooth out some
singularities. But serving as an ultra-violet cut-off higher
curvature terms could only possibly smooth out a real singularity
at $r<\epsilon$, not at $r>\epsilon$. And this is precisely what
happens in our solution - the presence of higher curvature terms
ensures that we have only a coordinate singularity at $r<\epsilon$
instead of a real naked singularity as would be the case had we
included only the Einstein-Hilbert term.

\section{Conclusions}

{}In the presence of the Einstein-Hilbert and Gauss-Bonnet terms
in the bulk action we have smooth infinite-volume solutions which
exist for a range of positive values of the brane tension (the
diluting property). These solutions, therefore, provide examples
of brane world scenarios where the brane world-volume can be flat
without any fine-tuning or presence of singularities.

{}We suspect (albeit we do not have a proof of
this statement) that even for generic higher curvature terms
solutions with the aforementioned properties should still exist.
In particular, we suspect that the fact that we found non-singular
solutions has to do with including {\em higher curvature} terms in
the bulk rather than with their particular (Gauss-Bonnet)
combination, which we have chosen to make computations tractable.

{}There are many interesting open questions to be addressed in
scenarios with infinite-volume extra dimensions. As was originally
pointed out in \cite{DGP,witten,DG}, these scenarios offer a new
arena for addressing the cosmological constant problem. And
addressing the aforementioned open questions definitely seems to
be worthwhile. We hope our results presented in this talk will
stimulate further developments in this field.

\section*{Acknowledgments}

A.I. would like to thank Stony Brook University for support.

\end{document}